\documentclass[prb,twocolumn,showpacs]{revtex4}

\usepackage{graphicx}

\newcommand{\ba}{\begin{eqnarray}}
\newcommand{\be}{\begin{equation}}
\newcommand{\ea}{\end{eqnarray}}
\newcommand{\ee}{\end{equation}}

\newcommand{\ignore}[1]{}

\begin{document}

\title{Reentrant magnetic ordering and percolation in a spin-crossover system}

\author{Carsten Timm}
\email{ctimm@ku.edu}
\affiliation{Department of Physics and Astronomy, University of Kansas,
Lawrence, Kansas 66045, USA}
\author{Charles J. Pye}
\affiliation{Department of Physics and Astronomy, University of Kansas,
Lawrence, Kansas 66045, USA}

\date{February 8, 2008}

\begin{abstract}
Spin-crossover compounds, which are characterized by magnetic ions showing
low-spin and high-spin states at thermally accessible energies, are ubiquitous
in nature. We here focus on the effect of an exchange interaction on the
collective properties for the case of non-magnetic low-spin ions, which applies
to $\mathrm{Fe}^{2+}$ compounds. Monte Carlo simulations are used to study a
three-dimensional spin-crossover model for the full parameter range from
essentially pure high spin to essentially pure low spin. We find that as the
low-spin state becomes more favorable, the Curie temperature drops, the
universality class deviates from the three-dimensional Heisenberg class, and
the transition eventually changes to first order. A heat-bath algorithm that
grows or shrinks low-spin and high-spin domains is developed to handle the
first-order transition. When the ground state has low spin, a reentrant
magnetic transition is found in a broad parameter range. We also observe a
percolation transition of the high spins, which branches off the first-order
magnetic transition.
\end{abstract}

\pacs{
75.10.Hk, 
75.20.Ck, 
75.50.Xx  
}

\maketitle

\section{Introduction}
\label{sec.intro}

Spin-crossover materials are characterized by ions or atoms that can be in
either a low-spin (LS) or a high-spin (HS) state. If the LS state is the ground
state, they often show a crossover or phase transition to HS at higher
temperatures, since the HS state has larger degeneracy and is thus entropically
favored.\cite{CaS31} Spin-crossover systems include organometallic
complexes,\cite{SCOrev,BlP04,JSR07} organic radicals,\cite{FuA99,ICC02}
Prussian Blue analogues,\cite{FMO95,SIF96,GRN00,BLE00} and other inorganic
transition-metal salts.\cite{Co,LVJ07} For example, a principal mineral in the
Earth's lower mantle is the spin-crossover compound
$\mathrm{Mg}_{1-x}\mathrm{Fe}^{}_x\mathrm{O}$.\cite{LVJ07}

If the spins are carried by transition-metal ions, the energy difference between
HS and LS is due to the interplay of crystal-field splitting of the
\textit{d}-levels and Hund's rule coupling.\cite{SCOrev,BlP04} For
$\mathrm{Fe}^{2+}$ ions the LS state has spin quantum number $S=0$, whereas the
HS state has $S=2$. Consequently, LS ions are essentially non-magnetic and
effects due to the LS/HS transition are expected to be pronounced. These systems
are related to diluted spin models,\cite{VMG02,dilute,YRH04,PWD07,GoR07} but in
spin-crossover systems the LS/HS degree of freedom is not quenched but
\emph{dynamical}.


Spin-crossover systems show interesting collective behavior, since they have two
coupled degrees of freedom: the Heisenberg-type spin orientation and the
Ising-type LS/HS degree of freedom. In one-dimensional systems, quantum
fluctuations are important, in particular in the case of an antiferromagnetic
exchange interaction.\cite{TiS05,Tim06} At zero temperature quantum effects in
this case stabilize complex phases with magnetic unit-cell lengths of
$3,5,7,\ldots$ lattice constants.\cite{TiS05}


In this paper, we study a three-dimensional (3D) spin-crossover model including
Heisenberg spins with isotropic exchange interaction. The LS state has spin
$S=0$. This is the 3D version of the spin-crossover chain studied by Timm and
Schollw\"ock.\cite{TiS05} We expect quantum fluctuations to be much weaker in
3D and employ a classical description. Nishino and coworkers \cite{NYM98} have
studied a 3D model with \emph{Ising}-type exchange, low spin $S=0$, and high
spin $S=1/2$, i.e., two possible orientations of the high spins, employing
mean-field theory and Monte Carlo (MC) simulations. Konishi \textit{et
al}.\cite{KTN06}\ have studied an Ising-type model for Prussian Blue analogues.
Our model is different in that we describe Heisenberg spins, which are 3D unit
vectors $\mathbf{S}_i$ interacting via an isotropic exchange interaction.

We perform MC simulations for the Heisenberg spin-crossover model on a simple
cubic lattice. We are interested in the overall phase diagram, not in
high-precision values for the transition temperature or critical exponents. In
Sec.~\ref{sec.sim} we briefly introduce the model and discuss the simulation
technique. In Sec.~\ref{sec.results} we present and discuss our results, which
are summarized in Sec.~\ref{sec.sum}.

\section{Model and simulations}
\label{sec.sim}


We are interested in 3D systems containing 
$\mathrm{Fe}^{2+}$ ions. Since the HS state has $S=2$ and the system is
3D, we expect quantum fluctuations to be relatively weak. We
therefore start from the classical Hamiltonian
\be
H = -V \sum_{\langle ij\rangle} \sigma_i \sigma_j
    -\Delta \sum_i \sigma_i\
    -J \sum_{\langle ij\rangle} \mathbf{S}_i \cdot \mathbf{S}_j ,
\label{H.1}
\ee
where $\sigma_i$ equals $+1$ ($-1$) and $|\mathbf{S}_i|$ is
zero (unity) for the LS (HS) state. We thus absorb the magnitude of the high
spins into the exchange coupling $J$. $i$ runs over all sites of a simple cubic
lattice, while $\langle ij\rangle$ denotes all nearest-neighbor bonds, counting
each bond once. We can map the
antiferromagnetic case $J<0$ onto the ferromagnetic case $J>0$ by flipping all
spins on one sublattice. We thus restrict ourselves to $J\ge0$ without loss of
generality and express all energies in units of $J$. The coupling $V$ describes
the affinity for equal LS/HS states on neighboring sites. This interaction is
thought to be of mostly elastic origin.\cite{elastic,NBY07} Finally, the Ising
``magnetic field'' $\Delta$ describes the on-site energy difference between HS
and LS states. To be precise, the energy difference is
$2\Delta$ and $\Delta > 0$ favors a LS ground state.


The ground state is a ferromagnetic HS state for $\Delta/J<3/2$ and
$V>\Delta/6-J/2$, a pure LS state for $\Delta/J>3/2$ and $V>-\Delta/6$, and a
checkerboard LS/HS state for $V<\Delta/6-J/2$ and $V<-\Delta/6$. This is
similar to the $T=0$ phase diagram for one-dimensional chains with
ferromagnetic Ising exchange interaction.\cite{Tim06} We mostly concentrate on
the case $V=0$.

We also need to specify the degeneracies of the LS and HS states. In the MC
simulation these govern the attempt rates for LS-to-HS and HS-to-LS flips. Only
the ratio of degeneracies is important because the physics is not affected by
an overall constant factor in the partition function. For HS $S=2$ the
underlying quantum system has a degeneracy factor of $g_{\mathrm{HS}}=5$ per
site in the HS state and $g_{\mathrm{LS}}=1$ in the LS state. However, it is
well known that the ratio $G\equiv g_{\mathrm{HS}}/g_{\mathrm{LS}}$ can be
strongly enhanced due to softer vibrations in the HS
state.\cite{BoK95,BCV95,YMK98,GKA00,NBY07}

Our results apply to arbitrary degeneracies since
one can absorb $G$ into an effective
Ising magnetic field\cite{WaP70,Don78,NBY07}
\be
\Delta_{\mathrm{eff}} \equiv \Delta - \frac{T}{2}\, \ln G
\label{Beff.1}
\ee
(we take $k_B=1$).
The simulations are therefore performed for $G=1$, i.e., with the same attempt
rates for LS-to-HS and HS-to-LS flips. For specific systems,
$\Delta$ is then replaced by $\Delta_{\mathrm{eff}}$.

We now turn to the MC simulations for this model. We simulate finite systems of
size $L\times L\times L$ with periodic boundary conditions. The Heisenberg spins
$\mathbf{S}_i$ are stored even for the LS sites, but the acceptance
probabilities do not depend on them. We employ several different updates: (a) We
use Wolff single-cluster updates\cite{Wol89} for the Heisenberg spins
$\mathbf{S}_i$, but restricted to percolating clusters of high spins. This means
that we choose a random site and, if it is in the HS state, grow a cluster to be
flipped as in the original algorithm,\cite{Wol89} but only attempting to add
neighboring HS sites. These cluster moves mostly avoid critical slowing down
close to the second-order magnetic-ordering transition.




(b) We employ cluster updates of the LS/HS degree of freedom, which are
also based on the Wolff algorithm\cite{Wol89} and use a ghost spin to describe
the local effective Ising magnetic field $\Delta$. The approach is similar to
R\"o\ss{}ler's.\cite{Ros99} Since the Heisenberg spins are not updated,
we absorb them into effective
parameters for the Ising model of the $\sigma_i$. We first make the LS/HS
dependence of the exchange term explicit by rewriting it as
\be
-J \sum_{\langle ij\rangle} \frac{1-\sigma_i}{2}\,\frac{1-\sigma_j}{2}\,
  \mathbf{S}_i\cdot\mathbf{S}_j .
\ee
With this notation we can use $\mathbf{S}_i$ with unit magnitude
regardless of whether a site is in the LS or HS state.
Apart from a term that is
independent of the $\sigma_i$, the Hamiltonian now reads
\be
\tilde H = - \sum_{\langle ij\rangle} \tilde V_{ij}\, \sigma_i \sigma_j
  - \sum_i \tilde \Delta_i\, \sigma_i
\label{IsingH.1}
\ee
with
\ba
\tilde V_{ij} & \equiv & V + \frac{J}{4}\, \mathbf{S}_i\cdot\mathbf{S}_j , \\
\tilde \Delta_i & \equiv & \Delta_{\mathrm{eff}} - \frac{J}{4}\, \mathbf{S}_i
  \cdot \!\!\! \sum_{\mathrm{NN}\: j\: \mathrm{of}\: i} \! \mathbf{S}_j ,
\ea
where the last sum is over the nearest neighbors of $i$.
The Hamiltonian (\ref{IsingH.1}) describes an Ising model with nonuniform
interaction in a nonuniform field. We rewrite the external field as the
interaction with a ghost spin $\sigma_g$, which couples to all sites $i$ with
coupling $\tilde V_{ig}$. If we choose $\sigma_g=1$, we require $\tilde
V_{ig}=\tilde \Delta_i$. The Hamiltonian then reads
\be
\tilde H = - {\sum_{\langle ij\rangle}}^\prime \tilde V_{ij}\, \sigma_i \sigma_j ,
\label{IsingH.2}
\ee
where the sum includes the ghost site.
In the simulation, we chose a random site and
try to add all bonds emanating from it. The probability for adding
a bond from $i$ to $j$ is $P_{ij} = \max(0, 1-e^{\Delta E/T})$, where $\Delta
E$ is the change in energy according to Eq.~(\ref{IsingH.2}). If
$j$ is a normal lattice site, we really add it, but if it is the ghost site, we
reject the entire cluster update. We then repeat these steps for all newly
added sites. If we fail to add more sites but have not rejected the update,
we flip all $\sigma_i$ on the cluster. This algorithm satisfies detailed
balance.\cite{CzR97,Ros99}


We also include (c) local spin rotations and (d) local LS/HS Ising spin flips
within the Metropolis algorithm as a fallback when the cluster updates are
inefficient. For a LS site, any spin rotation is accepted, since it does not
cost any energy. We attempt approximately equal numbers of MC sweeps using
(a)--(d), where one sweep is defined as touching every spin once on average.

The resulting routine is very robust, but suffers from one problem: In a
relevant parameter range the magnetic transition is strongly first order.
Standard methods to deal with first-order transitions such as multicanonical
algorithms\cite{BeN91,Mar98} or entropic sampling\cite{Lee93,StH98} fail for an
interesting reason: The low-temperature phase has a high HS fraction and
ferromagnetic order, whereas the high-temperature phase shows a lower HS
fraction and no magnetic order. In the high-temperature phase the Heisenberg
spins are decoupled and typically randomly oriented. Starting from this phase,
local LS/HS updates are inefficient because the HS phase is stabilized by the
exchange interaction, which is zero for isolated high spins. However, LS/HS
cluster updates also have very low probability of reaching the HS phase, since
the random Heisenberg spins are energetically very unfavorable in the HS state.
Thus we need updates that create large clusters of high spins \emph{and} align
them. It is difficult to do this efficiently while satisfying detailed balance.


Our algorithm uses a heat-bath technique that attempts to insert a layer of high
spins which are well aligned with their neighbors. In detail, at any temperature
we start by simulating two replicas of the system, which only differ in the
starting configuration, which is pure HS with perfect ferromagnetic order and
pure LS with random $\mathbf{S}_i$, respectively. If both phases are (meta-)
stable and not very close together in configuration space, these replicas reach
quasi-equilibrium in each phase, but do not overcome the barrier between them.
We then take half of each replica and glue them together, thereby creating two
phase boundaries. For reasons discussed below we cut in planes perpendicular to
the $(111)$ direction.

The resulting system is then equilibrated using the updates (a)--(d) and
one additional type (e) consisting of randomly selecting two layers $L_1$ and
$L_2$ that are perpendicular to $(111)$, removing $L_1$, and shifting all
layers between $L_1$ and $L_2$ (inclusive) by one primitive lattice vector
{\boldmath$\hat x$}. We then create new spins in layer $L_2$ using a heat-bath
algorithm: At each site $i\in L_2$, $\sigma_i$ and $\mathbf{S}_i$ are randomly
chosen with the proper Boltzmann probability distribution.
The reason for choosing planes
perpendicular to $(111)$ is that each site in $L_2$ only has neighbors that are
not in $L_2$. Thus we can select each $(\sigma_i,\mathbf{S}_i)$ independently.
If we had chosen layers perpendicular to $(100)$ we would have had to create a
Boltzmann distribution of the entire layer.
During the equilibration of the glued-together system, the updates (e) grow the
more favorable phase and eventually remove the less favorable one.


Finally, we perform measurements for the equilibrated system. We wait for at
least one MC sweep between measurements. Since the resulting time series is
correlated, the errors are estimated using the blocking
method.\cite{FlP89} The most important quantities we measure are the
average square of the magnetization,
\be
\langle\mathbf{M}^2\rangle \equiv \bigg\langle \frac{1}{L^3}
\sum_{\mathrm{HS}\:i} \mathbf{S}_i \cdot \frac{1}{L^3}
\sum_{\mathrm{HS}\:j} \mathbf{S}_j \bigg\rangle ,
\ee
where the sums are only over HS sites, the average fourth power of the
magnetization, $\langle \mathbf{M}^4\rangle$, and the HS fraction
\be
\gamma \equiv \bigg\langle \frac{1}{L^3} \sum_i \frac{1-\sigma_i}{2}
  \bigg\rangle .
\ee
Note that $\sigma_i=1$ ($-1$) for LS (HS).

The Binder cumulant\cite{Bin81}
\be
C_2 \equiv \frac{5}{2}
  - \frac{3}{2}\, \frac{\langle \mathbf{M}^4\rangle}
  {\langle \mathbf{M}^2\rangle^2}
\label{Binder.1}
\ee
is not as useful for the precision determination
of the Curie temperature $T_C$ as for the pure
Heisenberg model, since our model has two length scales, the correlation
lengths $\xi$ of the Heisenberg spins $\mathbf{S}_i$ and $\xi_\sigma$ of the
LS/HS degree of freedom $\sigma_i$. We expect $\xi_\sigma$ to be a continuous
function of temperature through the magnetic transition like the HS fraction
$\gamma$, as long as the transition is of second order. The Binder cumulant can
be written as a scaling function $C_2 = C_2(t L^{1/\nu}, L/\xi_\sigma)$,
where $t\equiv (T-T_C)/T_C$ and $\nu$ is the critical exponent of $\xi$. At
$T=T_C$, we have $C_2 = C_2(0,L/\xi_\sigma(T_C))$ so that the Binder
cumulant does not become independent of system size. However, it is
still true that the Binder cumulant approaches a step function for $L\to\infty$.
For our definition (\ref{Binder.1}) the step is from unity in the ordered phase
to zero in the disordered phase. We can thus use $C_2=1/2$ as a criterion for
$T_C$ that is correct in the limit of large $L$.

We also measure the fraction $P_z$ of configurations with a site-percolating HS
cluster that wraps around the periodic boundaries in the \textit{z}-direction,
where the presence of this cluster is determined by a variant of the
algorithm of Machta \textit{et al}.\cite{MCL96} $P_z(T)$ curves for different
system sizes $L$ are expected to intersect at the percolation transition in the
limit of large $L$. Other definitions of the wrapping probability, e.g.,
wrapping in all directions, give the same result for the
transition.\cite{NeZ01}


\section{Results and discussion}
\label{sec.results}

In certain limiting cases, our model becomes either a pure Heisenberg model or
a pure Ising model. We first discuss the Ising limit. $\Delta$ is a constant
external field coupling to the Ising degree of freedom $\sigma_i$ and if we for
the moment assume the Heisenberg spins to be frozen, they contribute
nonuniform magnetic-field and coupling terms. The total Ising magnetic field
is nonzero unless $\Delta=J=0$ and breaks the symmetry under
$\sigma_i\to-\sigma_i$ and thus destroys the Ising critical point. The fact
that the Ising magnetic field depends on the $\mathbf{S}_i$ and thus fluctuates
does of course not restore the critical point.

Now consider the Heisenberg limit. The coupling to the LS/HS degree of freedom
$\sigma_i$ does not break spin-rotation symmetry. If we freeze the $\sigma_i$
we have a diluted Heisenberg model, as noted above. The Harris criterion
\cite{Har74} states that quenched disorder should be irrelevant for the
critical behavior if the critical exponent $\nu$ of the correlation length
satifies $\nu \ge 2/d$, where $d$ is the dimensionality of the system. For the
pure Heisenberg model, $\nu = 0.7112(5)$,\cite{CHP02} larger than
$2/3$. Since disorder is thus irrelevant, the Heisenberg critical behavior
should survive for a diluted system if it is not preempted by a first-order
transition. This is indeed found in Ref.~\onlinecite{GoR07}.

However, the $\sigma_i$ are \emph{dynamical}. Thus the Harris criterion does not
apply and different critical behavior is possible---we return to this point
below. Since spin-ro\-ta\-tion symmetry is not broken explicitly by the coupling
to $\sigma_i$, a spontaneous magnetic-ordering transition is still possible,
though.

In addition, there can be a site-percolation transition of the high
spins. Clearly, HS percolation is necessary but not sufficient for long-range
magnetic order.


\begin{figure}[t]
\centerline{\includegraphics[width=3.40in,clip]{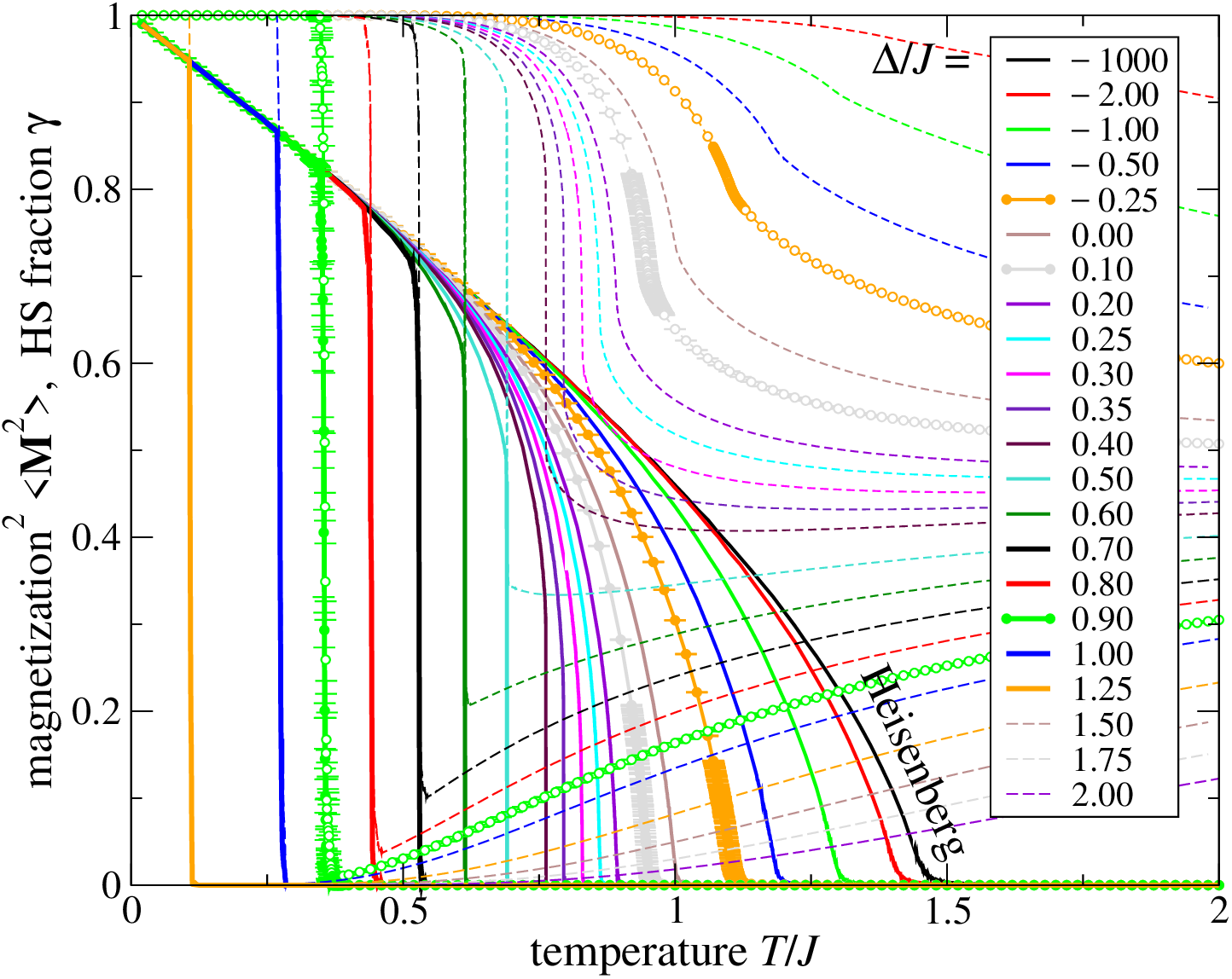}}
\caption{(Color online) Magnetization squared (solid curves) and HS fraction
(dashed curves) for elastic interaction $V=0$ and various values of $\Delta/J$.
For the magnetization curves, $\Delta$ increases from right to left. For
$\Delta/J\ge 3/2$ the magnetization vanishes at all temperatures. The results
have been obtained for system size $L=30$, $2000$ equilibration sweeps and at
least $10000$ measurement sweeps. For several values of $\Delta/J$ the data
points with error bars are shown.}
\label{fig.MofT.1}
\end{figure}

To prepare for the discussion of the phase diagram, we plot in
Fig.~\ref{fig.MofT.1} the magnetization squared, $\langle\mathbf{M}^2\rangle$,
and the HS fraction $\gamma$ as functions of temperature $T$ for elastic
interaction $V=0$ and various values of $\Delta/J$. Recall that $2\Delta$ is
the bare energy difference between HS and LS states. We find four regimes:

For $\Delta/J\alt 0.30$, the magnetization vanishes at a second-order
transition. The disordered phase above $T_C$ is predominantly HS.
$\Delta\to-\infty$ corresponds to the pure Heisenberg model, where our results
are consistent with $T_C/J \approx 1.457219(4)$ from high-precision
MC simulations.\cite{CHP02} $T_C$ decreases with increasing $\Delta$.

For $0.30 \alt \Delta/J \alt 0.55$, the magnetization vanishes at a
\emph{first}-order transition. The disordered phase is still mostly HS.

For $0.55 \alt \Delta/J < 3/2$, the transition is still of first order. The
disordered phase immediately above $T_C$ is now mostly LS, but the HS fraction
increases again with increasing temperature $T>T_C$.

For $\Delta/J \ge 3/2$ (exact value), there is no magnetic order at any
temperature. The system is in a pure LS state at $T=0$ and the HS fraction
increases smoothly with increasing temperature.

\begin{figure}[t]
\centerline{\includegraphics[width=3.40in,clip]{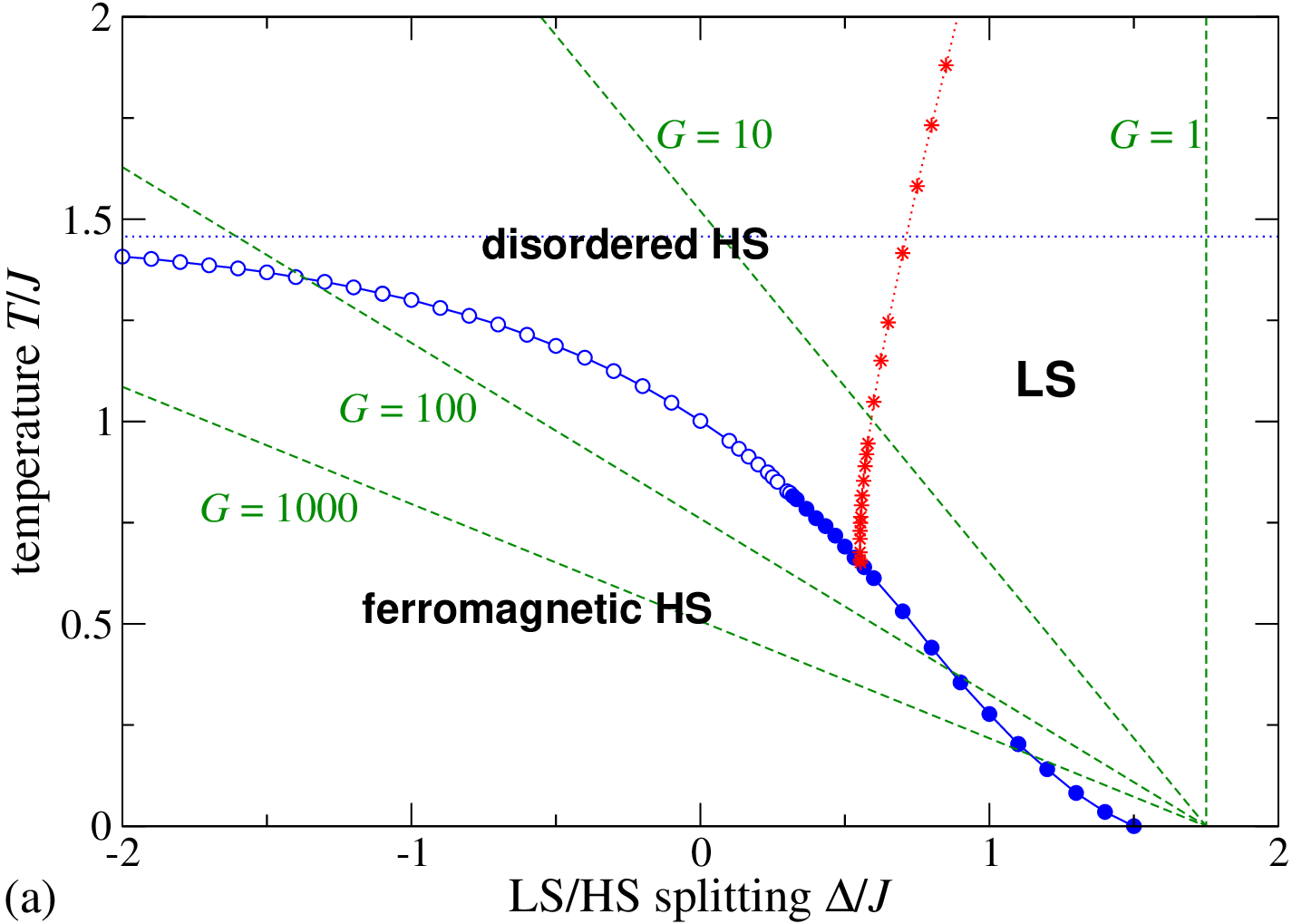}}
\centerline{\includegraphics[width=3.40in,clip]{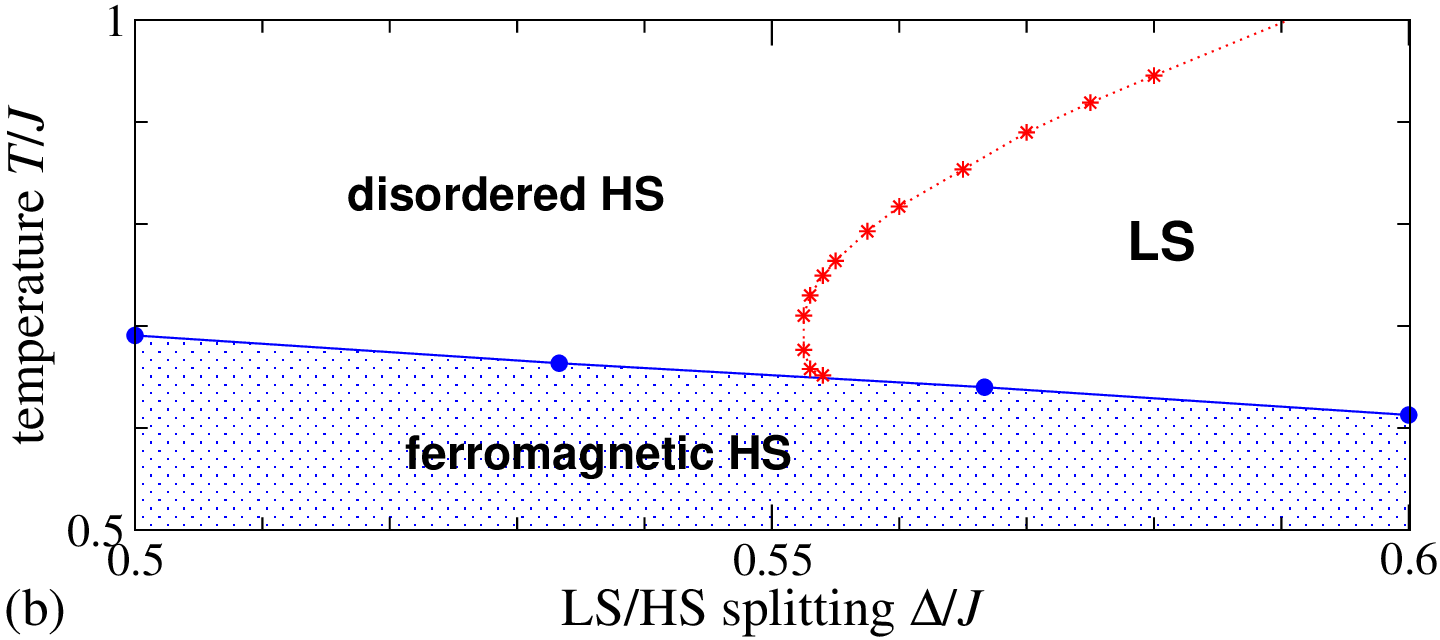}}
\caption{(Color online) (a)
Phase diagram of the 3D Heisenberg spin-crossover model
for elastic interaction $V=0$. The variables are (half) the bare LS/HS
splitting in units of the exchange coupling, $\Delta/J$, and temperature. The
solid curve with circles shows the magnetic transition. It approaches the pure
Heisenberg limit (dotted line) for $\Delta\to-\infty$. Open (filled) circles
represent a second-order (first-order) transition. The dotted curve with stars
shows the HS percolation transition. Also shown are typical lines of constant
$\Delta$ for ratios $G=g_{\mathrm{HS}}/g_{\mathrm{LS}}=1,10,100,1000$ of
effective degeneracies for $\Delta/J=1.75$. For larger $G$, reentrant magnetic
transitions are obvious. (b) Closeup of the intersection of magnetic and
percolation transitions in (a).}
\label{fig.PD.1}
\end{figure}

Figure \ref{fig.PD.1} shows the phase diagram in the $(\Delta,T)$ plane for
$V=0$. Since we are not interested in high-precision determination of transition
temperatures, we employ the following criteria to map out the transitions: $T_C$
is estimated from the Binder commulant $C_2=1/2$ found by bisection for $L=30$
with $20000$ equilibration sweeps and $2^{17}=131072$ measurement sweeps for
each data point. (For $\Delta/J=0.25$ the resulting $T_C$ is indistinguishable
on the scale of Fig.~\ref{fig.PD.1}(a) from the one obtained below from
finite-size scaling.) The temperature of the percolation transition is estimated
from the intersection of $P_z(T)$ for $L=30$ and $L=40$ found by bisection with
$20000$ equilibration sweeps and $131072$ measurement sweeps.

The phase diagram Fig.~\ref{fig.PD.1}(a) shows that $T_C$ decreases smoothly
with increasing $\Delta$. Note that $T_C$ is already significantly reduced for
$\Delta \lesssim 0$, where the ground state of an uncoupled site is still HS.
The low-temperature phase remains ferromagnetically ordered for small positive
$\Delta$ due to the exchange interaction, which favors HS. The character of the
magnetic transition changes from second order to first order at $\Delta/J =
0.313(5)$ and $T_C$ approaches zero continuously at $\Delta/J=3/2$.

The HS percolation transition intersects with the magnetic transition where the
latter is of first order. It then coincides with the first-order transition
down to the point $\Delta/J=3/2$, $T_C=0$. The intersection point does not have
any special properties; the percolation transition continues for the
undercooled magnetically disordered state (not shown). It is seen that on
approaching the magnetic transition, the percolation transition curves away
towards the LS phase. This shows that the HS phase is favored even though there
is no ferromagnetic order yet. This can be attributed to short-range
ferromagnetic correlations. The fluctuations responsible for these correlations
are relatively strong although the magnetic transition is of first order, since
the parameters are close to the endpoint of the first-order line. Figure
\ref{fig.PD.1}(b) shows that the percolation transition is reentrant as a
function of temperature in the vicinity of $\Delta/J=0.553$.


Recall that arbitrary ratios $G=g_{\mathrm{HS}}/g_{\mathrm{LS}}$ of effective
degeneracies can be incorporated by replacing $\Delta$ by
$\Delta_{\mathrm{eff}} = \Delta - (T/2)\, \ln G$. Curves of fixed $\Delta$ in
the $(\Delta_{\mathrm{eff}},T)$ phase diagram are straight lines of slope
$-2/\ln G$. Since $G>1$ (and often $G\gg 1$, Ref.~\onlinecite{BCV95}), the
slope is negative. Examples are shown in Fig.~\ref{fig.PD.1}(a). It is clear
that for not too small $G$ this leads to reentrant magnetic transitions: With
increasing temperature the system goes from paramagnetic (usually LS), to
ferromagnetic HS, to paramagnetic (usually HS). What happens physically is that
the increasing temperature entropically favors HS until the free energy of the
ferromagnetic HS state becomes lower than that of the LS state. At higher
temperature thermal fluctuations eventually destroy magnetic order, but HS
remains entropically favored.



\begin{figure}[t]
\centerline{\includegraphics[width=3.40in,clip]{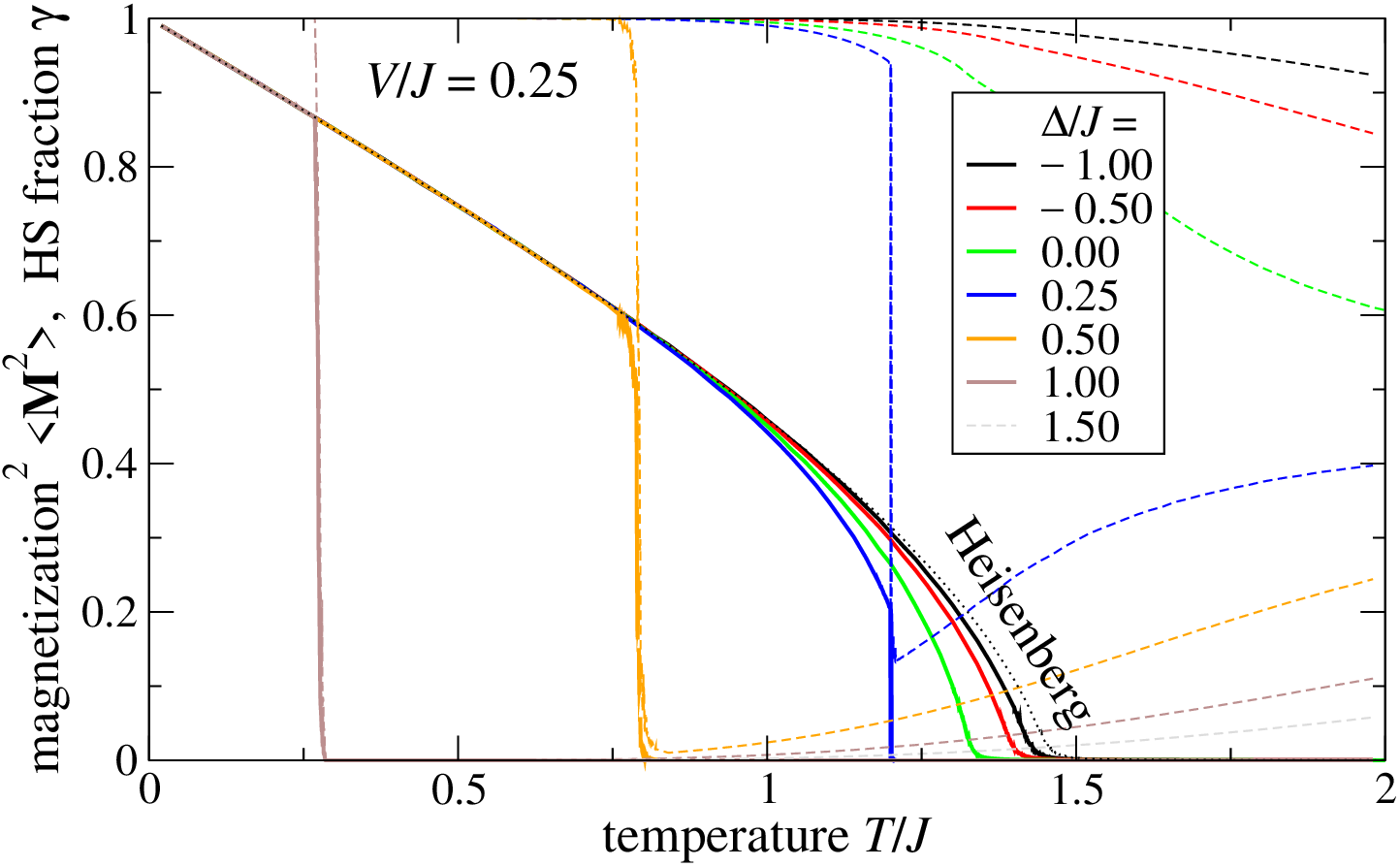}}
\caption{(Color online) Magnetization squared (solid curves) and HS fraction
(dashed curves) for elastic interaction $V/J=0.25$ and various values
of $\Delta/J$. For the magnetization curves, $\Delta$ increases from right
to left. For $\Delta/J\ge 3/2$ the magnetization vanishes at all
temperatures. The results have been obtained for system size $L=30$, $2000$
equilibration sweeps and at least $10000$ measurement sweeps.
The dotted curve shows $\langle\mathbf{M}^2\rangle$ for the Heisenberg case.}
\label{fig.MofT.2}
\end{figure}

We briefly consider the case of $V\neq 0$. Figure \ref{fig.MofT.2} shows the
magnetization squared and the HS fraction for $V/J=0.25$. Positive $V$ favors
neighboring sites both in the LS or both in the HS state, thereby
stabilizing phases with HS fractions $\gamma\approx 0$ or $\gamma\approx 1$. We
expect this to favor first-order magnetic transitions accompanied by a large
jump in $\gamma$, which is indeed seen in Fig.~\ref{fig.MofT.2}. For example,
for $\Delta/J=0.25$ the transition is now of first order, whereas it is of
second order for $V=0$, cf.\ Fig.~\ref{fig.MofT.1}. In addition, at the same
$\Delta/J$, positive $V$ stabilizes the ferromagnetic HS phase, increasing
$T_C$. However, the $T=0$ transition is at $\Delta/J=3/2$ for all $V/J\ge
-1/4$.

\begin{figure}[t]
\includegraphics[width=3.40in,clip]{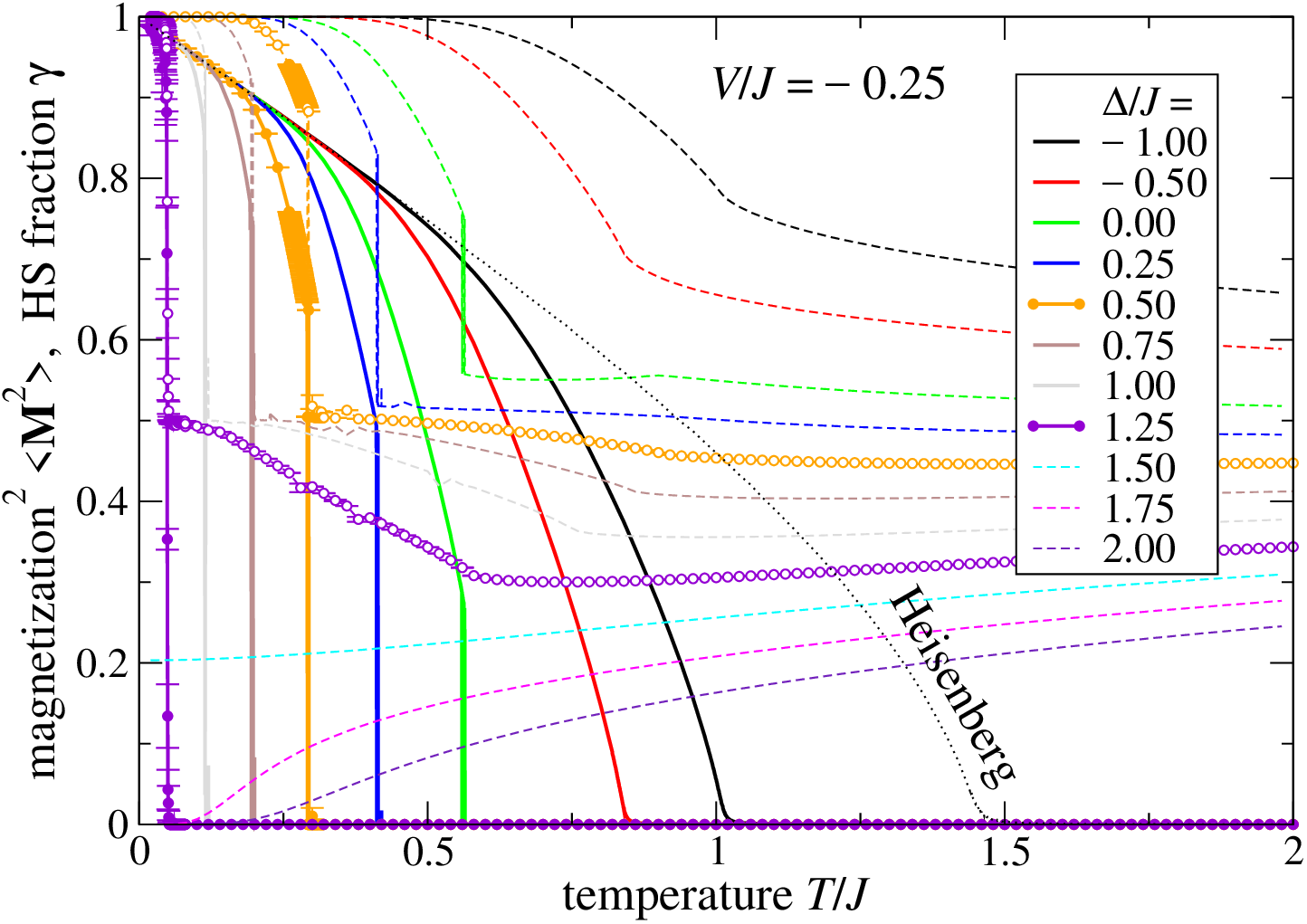}
\caption{(Color online) Magnetization squared (solid curves) and HS fraction
(dashed curves) for elastic interaction $V/J=-0.25$ and various values
of $\Delta/J$. For the magnetization curves, $\Delta$ increases from right to
left. For $\Delta/J\ge 3/2$ the magnetization vanishes at all
temperatures. This is a special point in the $T=0$ phase diagram
where ferromagnetic HS, pure LS, and checkerboard phases meet.
The results have been obtained for system size $L=30$, at least $2000$
equilibration sweeps and at least $10000$ measurement sweeps. For two values
of $\Delta/J$ the data points with error bars are shown. The dotted curve shows
$\langle\mathbf{M}^2\rangle$ for the Heisenberg case.}
\label{fig.MofT.3}
\end{figure}

Figure \ref{fig.MofT.3} shows the magnetization squared and the HS fraction for
$V/J=-0.25$. Negative $V$ favors neighboring sites in \emph{different} LS
and HS states. In fact for $V/J<-1/4$ the $T=0$ phase diagram contains a
checkerboard LS/HS phase at $\Delta/J$ around $3/2$. At the point $V/J=-1/4$,
$\Delta/J=3/2$ three phases (ferromagnetic HS, pure LS, and checkerboard) meet.

We find two phase transitions for intermediate values of $\Delta/J$: The HS
fraction $\gamma$ shows a sharp kink above $T_C$. We also note that the HS
fraction $\gamma$ is close to $1/2$ for $T\gtrsim T_C$, consistent with
checkerboard order of the LS/HS degree of freedom. An analysis of the average
staggered Ising order parameter, $\langle|L^{-3}\, \sum_i (-1)^i
\sigma_i|\rangle$ (not shown) finds long-range checkerboard order in the
intervening phase. This phase thus appears at finite temperatures for
$V/J=-0.25$, although at $T=0$ it is absent except at $\Delta/J=3/2$, as noted.


In addition, first-order transitions are again favored compared to $V=0$: For
$\Delta/J=0.25$ the transition is of first order for $V/J=-0.25$ and $0.25$ but
not for $V=0$. Furthermore, $T_C$ for the same $\Delta/J$ is
significantly reduced compared to $V\ge 0$, since neighboring LS/HS pairs are
of course unfavorable for magnetic order.


\begin{figure}[t]
\centerline{\includegraphics[width=3.40in,clip]{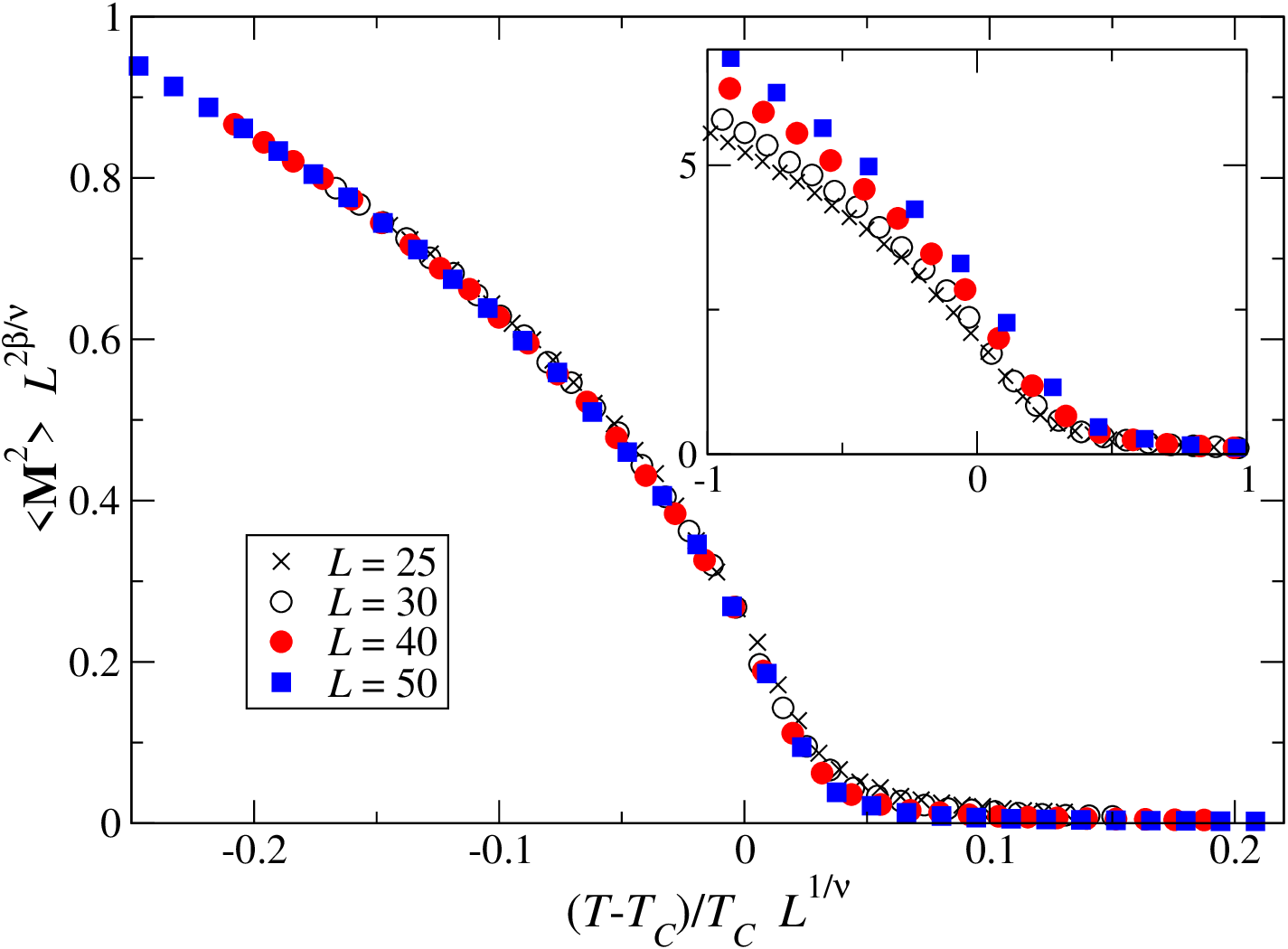}}
\caption{(Color online) Scaled magnetization squared,
$\langle\mathbf{M}^2\rangle\,
L^{2\beta/\nu}$, vs.\ scaled temperature, $t\,L^{1/\nu}$, for $V=0$,
$\Delta/J=0.25$ for system sizes $L=25,30,40,50$. The results have been obtained
using at least $2000$ equilibration sweeps and $262144$ measurement sweeps.
Inset: The same quantities with the same value of $T_C$
scaled with the exponents $\beta$, $\nu$ of the pure Heisenberg model.}
\label{fig.B025.scale}
\end{figure}

We next discuss the nature of the magnetic transition in some more detail. As
noted above, the Harris criterion\cite{Har74} is not applicable to our model.
It is thus reasonable to ask whether the \emph{dynamical} dilution changes the
universality class where the transition is of second order. To answer this
question, we perform finite-size scaling for the case of $\Delta/J=0.25$, for
which the transition is of second order, but close to the end point of the
first-order line, see Figs.~\ref{fig.MofT.1} and \ref{fig.PD.1}(a). The
magnetization squared close to the transition should scale like $\langle
\mathbf{M}^2\rangle\, L^{2\beta/\nu} \sim \Phi\!\left(t \, L^{1/\nu}\right)$
with $t\equiv (T-T_C)/T_C$, where $\Phi$ is a universal function. A
least-square fit for system sizes $L=40$ and $L=50$ yields $T_C/J = 0.8604(2)$,
$\beta=0.26(2)$, and $\nu=1.30(8)$. The errors are the ones incurred by the
least-square fit, which are much larger than the statistical errors of $\langle
\mathbf{M}^2\rangle$. The scaled magnetization squared is plotted vs.\ the
scaled temperature for sizes $L=25,30,40,50$ in Fig.~\ref{fig.B025.scale}. It
is apparent that corrections to scaling are relatively large. Nevertheless, the
behavior is clearly inconsistent with the 3D Heisenberg exponents
$\beta=0.3689(3)$ and $\nu=0.7112(5)$.\cite{CHP02} The inset in
Fig.~\ref{fig.B025.scale} shows that the scaling fails for Heisenberg critical
exponents. Since Figs.~\ref{fig.MofT.1} and \ref{fig.PD.1} do not show any
qualitative change in the second-order magnetic transition as a function of
$\Delta$, we conjecture that the critical exponents smoothly approach the 3D
Heisenberg limit for $\Delta\to-\infty$.


\begin{figure}[t]
\centerline{\includegraphics[width=3.40in,clip]{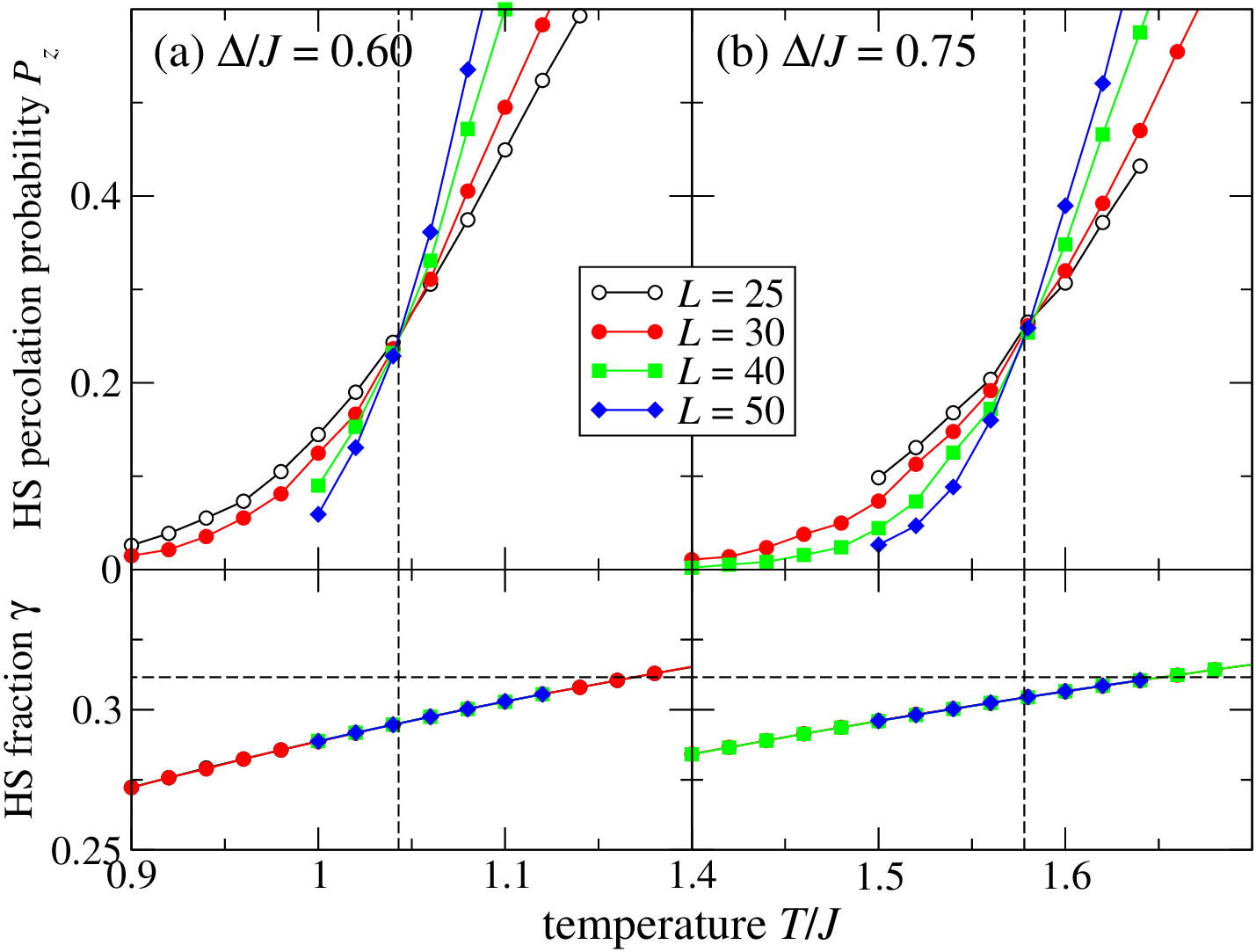}}
\caption{(Color online) Probabilities for finding a spanning HS cluster in the
\textit{z}-direction (upper panels) and HS fraction (lower panels) as functions
of temperature for (a) $\Delta/J=0.60$ and (b) $\Delta/J=0.75$ and system sizes
$L=25,30,40,50$. The horizontal dashed line indicates the critical
percolating HS
fraction in the absence of correlations. The vertical dashed lines mark the
positions of the percolation transitions.}
\label{fig.perc}
\end{figure}

Finally, we turn to the HS percolation transition. For uncorrelated $\sigma_i$,
the transition would take place at a HS fraction of $\gamma = \gamma_{p0} =
0.3116081(21)$.\cite{BFM99} In our case, there are short-range correlations
between the $\sigma_i$, even for $V=0$, due to the exchange interaction $J$,
which favors high spins at nearest-neighbor sites. We have seen that these
correlations shift the percolation transition towards larger $\Delta$. Thus
there is no reason to expect the HS percolation transition to take place at
$\gamma=\gamma_{p0}$. Figure \ref{fig.perc} shows the probability $P_z$ of
finding a HS cluster spanning the system in the \textit{z}-direction as a
function of temperature for two values of $\Delta$ and several system sizes.
The crossing of $P_z$ curves indicates the percolation transition.\cite{MCL96}
The lower panels show the HS fraction $\gamma$ and also the value
$\gamma_{p0}$. The HS fraction at the percolation transition, $\gamma_p$, is
inconsistent with $\gamma_{p0}$: For $\Delta/J=0.75$ we find $\gamma_p \approx
0.304$ and for $\Delta/J=0.6$, closer to $T_C$, $\gamma_p \approx 0.295$. Thus
the high spins percolate at a \emph{lower} HS fraction than they would for a
random distribution, as expected for positive correlations between neighboring
high spins.

\section{Summary}
\label{sec.sum}

We have performed MC simulations for a 3D spin-cross\-over model with Heisenberg
exchange interaction and spin $S=0$ in the LS state, as realized in
$\mathrm{Fe}^{2+}$ compounds. A full range of onsite energy differences
$2\Delta$ between HS and LS has been explored, from values strongly favoring HS
to values strongly favoring LS. We have focused on the case of negligible
elastic interaction. The main results are the following: The Curie temperature
is significantly reduced already on the HS side ($\Delta<0$), but initially
remains nonzero on the LS side ($\Delta>0$) due to the exchange interaction $J$,
which favors HS. As $\Delta$ increases further, the character of the transition
changes from second to first order and $T_C$ decreases continuously until it
reaches zero at a specific value of $\Delta$. A MC method involving growing and
shrinking LS and HS domains has been introduced to tackle the first-order
transition.


Where the magnetic transition is of second order, it is not in the 3D
Heisenberg universality class, at least close to the onset of first-order
transitions. The system also shows a HS site-percolation transition, which
intersects the first-order magnetic transition. Its position is affected by
short-range magnetic correlations. They also lead to a smaller HS fraction
$\gamma$ at the percolation transition than for a random percolation model.

For the realistic case of a LS ground state and an excited HS state with much
larger degeneracy, we find reentrant magnetic transitions. Increasing
temperature entropically favors HS, which stabilizes ferromagnetic order in an
intermediate temperature range.


\acknowledgments

We would like to thank Donald Priour, Jr.\ and Patrik Henelius for helpful
discussions.

\end{document}